\documentstyle[aps,floats,eqsecnum,psfig]{revtex}

\begin{document}
\draft

\title{
What can we learn about GW Physics with an elastic spherical antenna?}

\author{J. Alberto Lobo}
\address{Departament de F\'\i sica Fonamental \\
         Universitat de Barcelona, Spain.}

\date{\today}
\date{5 December 1994) \\ (Revised \today}

\maketitle

\begin{abstract}

A general formalism is set up to analyse the response of an
{\it arbitrary\/} solid elastic body to an {\it arbitrary metric\/}
Gravitational Wave perturbation, which fully displays the details of
the interaction antenna-wave. The formalism is applied to the spherical
detector, whose sensitivity parameters are thereby scrutinised. A
{\it multimode\/} transfer function is defined to study the amplitude
sensitivity, and absorption cross sections are calculated for a general
metric theory of GW physics. Their {\it scaling\/} properties are shown
to be {\it independent\/} of the underlying theory, with interesting
consequences for future detector design. The GW incidence direction
deconvolution problem is also discussed, always within the context
of a general metric theory of the gravitational field.

\end{abstract}

\section{Introduction}

The idea of building ultracryogenic spherical Gravitational Wave (GW)
antennae seems to be progressively winning adepts, even despite the
technological difficulties of various kinds posed by a project like that,
which every expert acknowledges. Confidence in its feasibility stems from
many years of experience: groups at Stanford, Louisiana State University,
Roma and Legnaro (Italy) and Perth (Western Australia) have constructed
and operated, at different levels, cryogenic cylindrical bars of the Weber
type \cite{ko92}. In particular, a {\it long term\/} strain sensitivity
$h=6 \times 10^{-19}$ for millisecond bursts has been reported from the
bar {\sl EXPLORER\/} \cite{as93}. The new generation ultracryogenic
cylinder {\sl NAUTILUS\/}, of the Frascati group \cite{as91}, is
beginning operation as these lines are written \cite{eupc}, with an
expected sensitivity nearly an order of magnitude better than the above.

Spherical antennae are considered by many to be the natural next step
in the development of resonant GW detectors
\cite{fo71,ad75,wp77,jm93,MZ94,clo}. The reasons for this new trend
essentially derive from the {\it improved sensitivity\/} of a sphere
---which can be nearly an order of magnitude better than a clyinder
having the same resonance frequency, see below and \cite{clo}---, and
from its {\it multimode capabilities\/}, first recognised by Forward
\cite{fo71} and further elaborated in \cite{wp77,jm93}.

Although some of the most relevant aspects of detector sensitivity
have already received attention in the literature, it seems to me
that a sufficiently general and flexible analysis of the interaction
detector-GW has not been satisfactorily developed to date. This
{\it theoretical\/} shortage has a number of {\it practical\/}
negative consequences, too. Traditional analysis, to mention but an
example, is almost invariably restricted to General Relativity or
scalar--tensor theories of gravity; while it may be argued that this
is already very general, any such argument is, as a matter of fact,
understating the potentialities actually offered by a spherical GW
antenna to help decide for or against any one specific theory of the
gravitational field on the basis of {\it experimental observation}.

I thus propose to develop in this paper a full fledged mathematical
formalism which will enable analysis of the antenna's response to a
completely general GW, i.e., making no {\it a priori\/} assumptions
about which is the correct theory underlying GW physics (other than,
indeed, that it is a {\it metric\/} theory), {\it and\/} also making
no assumptions about detector shape, structure or boundary conditions.
Considering things in such generality is not only ``theoretically nice''
---it also brings about new results and a {\it better understanding\/}
of older ones. For example, it will be proved that the sphere is the
{\it most efficient\/} GW elastic detector shape, and that higher mode
absorption cross sections {\it scale independently of GW physics}.
I will also discuss the direction of incidence deconvolution problem
in the context of a general metric theory of gravity.

The paper is organised as follows: section 2 is devoted to the development
of the general mathematical framework, leading to a formula in which an
elastic solid's response is related to the action of an arbitrary metric
GW impinging on it. In section 3 the general equations are applied to the
homogeneous spherical body, and a discussion of the {\it deconvolution\/}
problem is presented as well. Section 4 contains the description of the
sphere's sensitivity parameters, specifically leading to the concept of
{\it multimode,} or vector, {\it transfer function}, and to an analysis
of the {\it absorption cross section\/} presented by this detector to a
passing by GW. Conclusions and prospects are summarised in section 5,
and two appendices are added which include mathematical derivations.

\section{General mathematical framework}

In the mathematical model, I shall be assuming that the antenna is a
solid elastic body which responds to GW perturbations according to the
equations of classical non-relativistic linear Elasticity Theory
\cite{ll70}. This is fully justified since, as stressed above, GW
induced displacements will be very small indeed, and the speed of
such displacements much smaller than that of light for any forseeable
frequencies. Although our primary interest is a spherical antenna, the
considerations which follow in the remainder of this section {\it have
general validity for arbitrarily shaped isotropic elastic solids}.

Let ${\bf u}({\bf x},t)$ be the displacement vector of the infinitesimal
mass element sitting at point {\bf x} relative to the solid's centre of
mass in its unperturbed state, whose density distribution in that state
is $\rho({\bf x})$. Let $\lambda$ and $\mu$ be the material's elastic
Lam\'e coefficients. If a volume force density ${\bf f}({\bf x},t)$ acts
on such solid, the displacement field ${\bf u}({\bf x},t)$ is the solution
to the system of partial differential equations \cite{ll70}

\begin{equation}
 \rho \frac{\partial^2 {\bf u}}{\partial t^2} - \mu\nabla^2 {\bf u} -
      (\lambda+\mu)\,\nabla(\nabla{\bf\cdot}{\bf u}) = {\bf f}({\bf x},t)
      \label{2.1}
\end{equation}

with the appropriate initial and boundary conditions. A summary of
notation and general results regarding the solution to that system
is briefly outlined in the ensuing subsection, as they are necessary
for the subsequent developments in this paper, and also in future work.

\subsection{Separable driving force}

For reasons which will become clear later on, we shall only be interested
in driving forces of the separable type

\begin{equation}
   {\bf f}({\bf x},t) = {\bf f}({\bf x})\,g(t)    \label{2.2}
\end{equation}

or, indeed, linear combinations thereof. The solution to (\ref{2.1})
does not require us to specify the precise boundary conditions on
${\bf u}({\bf x},t)$ at this stage, but we need to set the initial
conditions. We adopt the following:

\begin{equation}
   {\bf u}({\bf x},0) = {\bf \dot{u}}({\bf x},0) = 0
\end{equation}

where \ ${\bf\dot{}}\equiv\partial/\partial t$, implying that the
antenna is at complete rest before observation begins at $t$=0. The
structure of the force field (\ref{2.2}) is such that the displacements
${\bf u}({\bf x},t)$ can be expressed by means of a {\it Green function\/}
integral of the form

\begin{equation}
   {\bf u}({\bf x},t) = \int_0^\infty{\bf S}({\bf x};t-t')\,g(t')\,dt'
   \label{2.6}
\end{equation}

The deductive procedure whereby ${\bf S}({\bf x};t-t')$ is calculated
can be found in many standard textbooks ---see e.g. \cite{tm87}. The
result is

\begin{equation}
  {\bf S}({\bf x};t) = \left\{\begin{array}{ll}
    0 & \ \ {\rm if}\ \ t\leq 0 \\[1 em]
    \sum_N\,\mbox{\large $\frac{f_N}{\omega_N}$}\,
    {\bf u}_N({\bf x})\sin\omega_Nt
    & \ \ {\rm if}\ \ t\geq 0 \end{array}\right.   \label{2.15}
\end{equation}

where

\begin{equation}
  f_N\equiv\frac{1}{M}\,\int_{\rm Solid}{\bf u}_{N}^*({\bf x})\cdot
  {\bf f}({\bf x})\,d^3x      \label{2.14}
\end{equation}

and ${\bf u}_N({\bf x})$ are the normalised {\it eigen--solutions\/} to

\begin{equation}
   \mu\nabla^2{\bf u}_N + (\lambda+\mu)\,\nabla(\nabla{\bf \cdot}{\bf u}_N)
      = - \omega_N^2\rho{\bf u}_N        \label{2.9}
\end{equation}

with suitable {\it boundary conditions}. Here $N\/$ represents an index,
or set of indices, labelling the {\it eigenmode\/} of frequency~$\omega_N$.
The normalisation condition is (arbitrarily) chosen so that

\begin{equation}
  \int_{\rm Solid}{\bf u}_{N'}^*({\bf x})\cdot{\bf u}_{N}({\bf x})
  \,\rho({\bf x})\,d^3x = M\,\delta_{N'N}     
  \label{2.10}
\end{equation}

where $M\/$ is the total mass of the solid, and the asterisk denotes
complex conjugation. Replacing now (\ref{2.15}) into (\ref{2.6}) we
can write the solution to our problem as a series expansion:

\begin{equation}
   {\bf u}({\bf x},t) =
   \sum_N\,\frac{f_N}{\omega_N}\,{\bf u}_N({\bf x})\,g_N (t)
   \label{2.16}
\end{equation}

where

\begin{equation}
   g_N(t)\equiv\int_0^t g(t')\,\sin\omega_N (t-t')\,dt'   \label{2.17}
\end{equation}

Equation (\ref{2.16}) is the {\it formal\/} solution to our problem; it
has the standard form of an orthogonal expansion and is valid for
{\it any\/} solid driven by a separable force like (\ref{2.2}) and
{\it any\/} boundary conditions. It is therefore {\it completely
general\/}, given that type of force.

Before we go on, it is perhaps interesting to quote a simple but useful
example. It is the case of a solid hit by a {\it hammer blow\/}, i.e.,
receiving a sudden stroke at a point on its surface. Exam of the response
of a GW antenna to such perturbation is being used for correct tuning and
monitoring of the device \cite{wjpc}. If the driving force density is
represented by the simple model 

\begin{equation}
   {\bf f}^{({\rm hb})}({\bf x},t) = {\bf f}_0\,
   \delta^{(3)}({\bf x}-{\bf x}_0)\,\delta(t)   \label{2.18}
\end{equation}

where ${\bf x}_0$ is the surface point hit, and ${\bf f}_0$ is a constant
vector, then the system's response is immediately seen to be

\begin{equation}
   {\bf u}^{({\rm hb})}({\bf x},t) =
   \sum_N\,\frac{f_N^0}{\omega_N}\,{\bf u}_N({\bf x})\sin\omega_N t
   \label{2.19}
\end{equation}

with $f_N^0=M^{-1}\,{\bf f}_0\!\cdot\!{\bf u}_N^*({\bf x}_0)$. A hammer
blow thus excites {\it all\/} the solid's normal modes, except those
{\it perpendicular\/} to ${\bf f}_0$, with amplitudes which are
{\it inversely proportional to the mode's frequency}. This is seen to
be a rather general result in the theory of sound waves in isotropic
elastic solids.

\subsection{The GW tidal forces}

An incoming GW manifests itself as a {\it tidal\/} force density; in the
long wavelength linear approximation \cite{mtw} it only depends on the
``electric'' components of the Riemann tensor:

\begin{equation}
   f_i({\bf x},t) = \rho c^2\,R_{0i0j}(t)\,x_j    \label{3.1}
\end{equation}

where $c\/$ is the speed of light, and sum over the repeated index {\it j\/}
is understood. In (\ref{3.1}) tidal forces are referred to the antenna's
centre of mass, and thus {\bf x} is a vector originating there. Note that
I have omitted any dependence of $R_{0i0j}$ on spatial coordinates, since
it only needs to be evaluated at the solid's centre. The Riemann tensor
is only required to first order at this stage \cite{we72}:

\begin{equation}
   R_{0i0j} = \frac{1}{2}\left(h_{ij,00}-h_{0i,0j}-h_{0j,0i}+h_{00,ij}
                         \right)   \label{3.2}
\end{equation}

where $h_{\mu\nu}$ are the perturbations to flat geometry\footnote{
Throughout this paper, {\it greek\/} indices ($\mu,\nu,\ldots$) will run
through space-time values 0,1,2,3; {\it latin\/} indices ($i,j,\ldots$)
will run through space values 1,2,3 only.}, always at the centre of mass
of the detector.

The form (\ref{3.1}) is seen to be a sum of three terms like (\ref{2.2})
---but this three term ``straightforward'' splitting is not the most
convenient, due to lack of invariance and symmetry. A better choice is
now outlined.

An {\it arbitrary symmetric\/} tensor ${\cal S}_{ij}$ admits the
following decomposition:

\begin{equation}
   {\cal S}_{ij}(t) = {\cal S}^{(S)}(t)\,E_{ij}^{(S)}\ +\ 
        \sum_{m=-2}^2\,{\cal S}^{(m)}(t)\,E_{ij}^{(m)}  \label{3.3}
\end{equation}

where $E_{ij}^{(m)}$ are 5 linearly independent {\it symmetric\/} and
{\it traceless\/} tensors, and $E_{ij}^{(S)}$ is a multiple of the
{\it unit\/} \mbox{tensor $\delta_{ij}$.} ${\cal S}^{(S)}(t)$ and
${\cal S}^{(m)}(t)$ are uniquely defined functions, whose explicit form
depends on the particular representation of the \mbox{$E\/$-matrices}
chosen. A convenient one is the following:

\begin{mathletters}
\label{3.4}
\begin{eqnarray}
  & E_{ij}^{(S)} = \left(\mbox{\large $\frac{1}{4\pi}$}
                 \right)^{\frac{1}{2}}\left(\begin{array}{ccc}
       1 & 0 & 0 \\ 0 & 1 & 0 \\ 0 & 0 & 1 \end{array}\right) & 
                 \label{3.4a}      \\*[1.2 em]
  & E_{ij}^{(0)} = \left(\mbox{\large $\frac{5}{16\pi}$}
                 \right)^{\frac{1}{2}}\left(\begin{array}{ccc}
      -1 & 0 & 0 \\ 0 & -1 & 0 \\ 0 & 0 & 2 \end{array}\right)\ ,\ \ 
  E_{ij}^{(\pm 1)} = \left(\mbox{\large $\frac{15}{32\pi}$}
                 \right)^{\frac{1}{2}}\left(\begin{array}{ccc}
       0 & 0 & \mp 1 \\ 0 & 0 & -i \\ \mp 1 & -i & 0 \end{array}\right)
             \ ,\ \ 
  E_{ij}^{(\pm 2)} = \left(\mbox{\large $\frac{15}{32\pi}$}
                 \right)^{\frac{1}{2}}\left(\begin{array}{ccc}
       1 & \pm i & 0 \\ \pm i & -1 & 0 \\ 0 & 0 & 0 \end{array}\right) &
                       \label{3.4b}
\end{eqnarray}
\end{mathletters}

The excellence of this representation stems from its ability to display
the {\it spin features\/} of the driving terms in (\ref{3.1}). Such
features are characterised by the relations

\begin{equation}
  E_{ij}^{(S)}\,n_i n_j = Y_{00}(\theta,\varphi)\ \ ,\ \ 
  E_{ij}^{(m)}\,n_i n_j = Y_{2m}(\theta,\varphi)  \label{3.6}
\end{equation}

where {\bf n}$\,\equiv\,${\bf x}/$|{\bf x}|$ is the radial unit vector,
and $Y_{lm}(\theta,\varphi)$ are spherical harmonics \cite{Ed60}. Details
about the above $E\/$-matrices are given in Appendix A. In particular,
the orthogonality relations (\ref{A.9}) can be used to invert (\ref{3.3}):

\begin{mathletters}
\label{3.77}
\begin{eqnarray}
   {\cal S}^{(S)}(t) & = & \frac{4\pi}{3}\,E_{ij}^{(S)}\,{\cal S}_{ij}(t)
                      \label{3.77a}  \\[1 em]
   {\cal S}^{(m)}(t) & = & \frac{8\pi}{15}\,E_{ij}^{*(m)}\,{\cal S}_{ij}(t)
                     \ ,\qquad m=-2,\ldots,2    \label{3.77b}
\end{eqnarray}
\end{mathletters}

where an asterisk denotes complex conjugation. Note that
${\cal S}^{(S)}(t) = \sqrt{4\pi}{\cal S}(t)/3$, where
${\cal S}(t)\equiv\delta_{ij}\,{\cal S}_{ij}(t)$ is the tensor's trace.

We now take advantage of (\ref{3.3}) to express the GW tidal force
(\ref{3.1}) as a sum of split terms like (\ref{2.2}):

\begin{equation}
   {\bf f}({\bf x},t) = {\bf f}^{(S)}({\bf x})\,g^{(S)}(t)\ +\ 
       \sum_{m=-2}^2\,{\bf f}^{(m)}({\bf x})\,g^{(m)}(t)      \label{3.7}
\end{equation}

with

\begin{mathletters}
\label{3.8}
\begin{eqnarray}
 f_i^{(S)}({\bf x}) = \rho\,E_{ij}^{(S)}\,x_j\ \ & , & \ \ \ 
 g^{(S)}(t) = \frac{4\pi}{3}\,E_{ij}^{*(S)}\,R_{0i0j}(t)\,c^2
 \label{3.8a} \\*[1 em]
 f_i^{(m)}({\bf x}) = \rho\,E_{ij}^{(m)}\,x_j\ \ & , & \ \ \ 
 g^{(m)}(t) = \frac{8\pi}{15}\,E_{ij}^{*(m)}\,R_{0i0j}(t)\,c^2
     \qquad (m = -2,\ldots,2)  \label{3.8b}
\end{eqnarray}
\end{mathletters}

Straightforward application of (\ref{2.16}) yields the formal solution
of the antenna response to a GW perturbation:

\begin{equation}
   {\bf u}({\bf x},t) = \sum_N\,\omega_N^{-1}{\bf u}_N({\bf x})\,\left[
   f_N^{(S)}\,g_N^{(S)}(t)\ +\ \sum_{m=-2}^2\,f_N^{(m)}\,g_N^{(m)}(t)
   \right]         \label{3.9}
\end{equation}

with the notation of (\ref{2.14}) and (\ref{2.17}) applied
{\it mutatis mutandi\/} to the terms in (\ref{3.8}).

Equation (\ref{3.9}) gives the response of an arbitrary elastic solid
to an incoming weak GW, {\it independently of the underlying gravity
theory\/}, be it General Relativity (GR) or indeed any other {\it metric\/}
theory of the gravitational interaction. It is also valid for
{\it any antenna shape\/} and {\it any boundary conditions}, thus giving
the formalism, in particular, the capability of being used to study the
response of a detector which is {\it suspended\/} by means of a mechanical
device in the laboratory site ---a situation of much practical importance.
It is therefore {\it very\/} general.

Equation (\ref{3.9}) also tells us that that only {\it monopole\/} and
{\it quadrupole detector modes\/} can possibly be excited by a metric GW.
The nice thing about (\ref{3.9}) is that it {\it fully\/} displays the
monopole-quadrupole {\it structure\/} of the solution to our fundamental
differential equations.

In a non-symmetric body, all (or {\it nearly\/} all) the modes have
monopole and quadrupole moments, and (\ref{3.9}) precisely shows how much
each of them contributes to the detector's response. A {\it homogeneous
spherical\/} antenna, which is very symmetric, has a set of vibrational
eigenmodes which are particularly well matched to the form (\ref{3.9}):
it only possesses {\it one\/} series of monopole modes and {\it one\/}
(five-fold degenerate) series of quadrupole modes ---see next section
and Appendix B for details. The existence of so {\it few\/} modes which
couple to GWs means that {\it all the absorbed incoming radiation energy
will be distributed amongst those few modes only}, thereby making the
sphere the {\it most efficient\/} detector, even from the sensitivity
point of view. The higher energy cross section {\it per unit mass\/}
reported for spheres on the basis of \mbox{GR \cite{clo},} for example,
finds here its qualitative explanation. The generality of (\ref{3.9}),
on the other hand, means that {\it this excellence of the spherical
detector is there independently of which is the correct GW theory}.

Before going further, let me mention another potentially useful
application of the formalism so far. Cylindrical antennas, for instance,
are usually studied in the {\it thin rod\/} approximation; although this
is generally quite satisfactory, equation (\ref{3.9}) offers the
possibility of eventually considering corrections to such simplifying
hypothesis by use of more realistic eigenfunctions, such as those given
in \cite{Ras,ric}. Recent new proposals for stumpy cylinder arrays
\cite{maf} may well benefit from the above approach, too.

\section{The spherical antenna}

To explore the consequences of (\ref{3.9}) in a particular case, the mode
amplitudes ${\bf u}_N({\bf x})$ and frequencies $\omega_N\/$ must be
specified. From now on I will focus on a {\it homogeneous sphere\/} whose
surface is free of tractions and/or tensions; the latter happens to be
quite a good approximation, even if the sphere is suspended in the static
gravitational field \cite{ol}.

The normal modes of the free sphere fall into two families: so called
{\it toroidal\/} ---where the sphere only undergoes twistings which
keep its shape unchanged throughout the volume--- and {\it spheroidal\/}
\cite{mzno}, where radial as well as tangential displacements take place.
I use the notation

\begin{equation}
  {\bf u}_{nlm}^T({\bf x})\,e^{\pm i\omega_{nl}^T t}\qquad ,\qquad
  {\bf u}_{nlm}^P({\bf x})\,e^{\pm i\omega_{nl}^P t}  \label{3.10}
\end{equation}

for them, respectively; note that the index $N\/$ of the previous section is
a {\it multiple\/} index $\{nlm\}$ for {\it each\/} family; $l\/$ and $m\/$
are the usual {\it multipole\/} indices, and $n\/$ numbers from 1 to
$\infty$ each of the $l\/$-pole modes. The frequencies happen to be
independent of $m$, and so every one mode (\ref{3.10}) is (2$l\/$+1)-fold
degenerate. Further details about these eigenmodes are given in Appendix B.

In order to see what (\ref{3.9}) looks like in this case, integrals of
the form (\ref{2.14}) ought to be evaluated. It is straightforward to
prove that they all vanish for the toroidal modes, the spheroidal modes
contributing the only non-vanishing terms; after some algebra one finds

\begin{mathletters}
\label{3.11}
\begin{eqnarray}
  f_{nlm}^{(S)} & \equiv & \frac{1}{M}\,\int_{\rm Sphere}
  {\bf u}_{nlm}^{P\,*}({\bf x})\cdot{\bf f}^{(S)}({\bf x})\,d^3x =
  a_n\,\delta_{l0}\,\delta_{m0}
               \label{3.11a}  \\*[1 em]
  f_{nlm}^{(m')} & \equiv & \frac{1}{M}\,\int_{\rm Sphere}
  {\bf u}_{nlm}^{P\,*}({\bf x})\cdot{\bf f}^{(m')}({\bf x})\,d^3x =
   b_n\,\delta_{l2}\,\delta_{m'm}
  \label{3.11b}
\end{eqnarray}
\end{mathletters}

where

\begin{mathletters}
\label{3.12}
\begin{eqnarray}
  a_n & = & -\frac{1}{M}\,\int_0^R A_{n0}(r)\,\rho\,r^3\,dr
  \label{3.12a}    \\*[1 em]
  b_n & = & -\frac{1}{M}\,\int_0^R \left[A_{n2}(r) +
  3\,B_{n2}(r)\right]\,\rho\,r^3\,dr         \label{3.12b}
\end{eqnarray}
\end{mathletters}

The functions $A_{nl}(r)$, $B_{nl}(r)$ are given in Appendix B, and $R\/$
is the sphere's radius. To our reassurance, only the monopole and quadrupole
{\it sphere modes\/} survive, as seen by the presence of the factors
$\delta_{l0}\/$ and $\delta_{l2}\/$ in (\ref{3.11a}) and (\ref{3.11b}),
respectively. The final series is thus a relatively simple one, even in
spite of its generality\footnote{
From now on I will drop the label $P\/$, meaning {\it spheroidal\/}
mode, to ease the notation since {\it toroidal\/} modes no longer appear in
the formulae.}:

\begin{equation}
   {\bf u}({\bf x},t) = \sum_{n=1}^\infty\,\frac{a_n}{\omega_{n0}}
   \,{\bf u}_{n00}({\bf x})\,g_{n0}^{(S)}(t)\ +\ 
   \sum_{n=1}^\infty\,\frac{b_n}{\omega_{n2}}\,\left[\sum_{m=-2}^2
   \,{\bf u}_{n2m}({\bf x})\,g_{n2}^{(m)}(t)\right]\ ,\qquad (t>0)
   \label{3.16}
\end{equation}

where, it is recalled,

\begin{equation}
 g_{nl}^{(S,m)}(t)=\int_0^t g^{(S,m)}(t')\,\sin\omega_{nl}(t-t')\,dt'
 \ ,\qquad (m=-2,\ldots,2)     \label{3.6b}
\end{equation}

Equation (\ref{3.16}) constitutes the sphere's response to an arbitrary
tidal GW perturbation, and will be used to analyse the sensitivity of the
spherical detector in the next section. Before doing so, however, a few
comments on the antenna's signal {\it deconvolution capabilities\/}, within
the context of a completely general metric theory of GWs, are in order.

\subsection{The deconvolution problem}

Let us first of all take the Fourier transform of (\ref{3.16}):

\begin{equation}
  {\bf U}({\bf x},\omega)\equiv\int_{-\infty}^\infty
  {\bf u}({\bf x},t)\,e^{-i\omega t}\,dt            \label{3.17}
\end{equation}

This is seen to be

\begin{eqnarray}
  {\bf U}({\bf x},\omega) & = & \frac{\pi}{i}\,\sum_{n=1}^\infty\,
  \frac{a_n}{\omega_{n0}}\,{\bf u}_{n00}({\bf x})\,G^{(S)}(\omega)
  \left[\delta(\omega\!-\!\omega_{n0}) - \delta(\omega\!+\!
  \omega_{n0})\right]
  \ + \nonumber \\*[1 em]
   & + & \!\frac{\pi}{i}\,\sum_{n=1}^\infty\,\frac{b_n}{\omega_{n2}}
   \left[\sum_{m=-2}^2\,{\bf u}_{n2m}({\bf x})\,G^{(m)}(\omega)\right]
   \left[\delta(\omega\!-\!\omega_{n2}) - \delta(\omega\!+\!
   \omega_{n2})\right]
   \label{3.18}
\end{eqnarray}

where $G^{(S)}(\omega)$ and $G^{(m)}(\omega)$ are the Fourier transforms
of $g^{(S)}(t)$ and $g^{(m)}(t)$, respectively:

\begin{equation}
  G^{(S,m)}(\omega)\equiv\int_0^\infty g^{(S,m)}(t)\,e^{-i\omega t}\,dt
  \label{3.19}
\end{equation}

The $\delta\/$-function factors are of course idealisations corresponding
to infinitely long integration times and infinitely narrow resonance
linewidths ---but the essentials of the ensuing discussion will not be
affected by those idealisations.

If the measuring system were (ideally) sensitive to all frequencies,
filters could be applied to examine the antenna's oscillations at each
monopole and quadrupole frequency: a single transducer would suffice to
reveal $G^{(S)}(\omega)$ around the monopole frequencies $\omega_{n0}$,
whilst {\it five\/} (placed at suitable positions) would be required to
calculate the five degenerate amplitudes $G^{(m)}(\omega)$ around the
quadrupole frequencies $\omega_{n2}$. Once the {\it six\/} functions
$G^{(S,m)}(\omega)$ would have thus been determined, inverse Fourier
transforms would give us the functions $g^{(S,m)}(t)$, and thereby the
six Riemann tensor components $R_{0i0j}(t)$ through inversion of the
second equations (\ref{3.8}), i.e., as an expansion like (\ref{3.3})
---only with $g\/$'s instead of ${\cal S}\/$'s. Deconvolution would
then be complete.

Well, not quite\ldots\ Knowledge of the Riemann tensor in the
{\it laboratory\/} frame coordinates is not really sufficient to say the
waveform has been completely deconvolved, unless we {\it also\/} know the
{\it source position\/} in the sky. There clearly are two possibilities:

\begin{enumerate}
\item[{\sf i)}] The source position {\it is\/} known ahead of time by some
other astronomical observation methods. Let me rush to emphasise that, far
from trivial or uninteresting, this is a {\it very important\/} case to
consider, specially during the first stages of GW Astronomy, when any
reported GW event will have to be thoroughly checked by all possible means.

If the incidence direction is known, then a rotation must be applied to
the just obtained quantities $R_{0i0j}(t)$, which takes the laboratory
\mbox{$z\/$-axis} into coincidence with the incoming wave propagation
vector. A classification procedure must thereafter be applied to the so
transformed Riemann tensor in order to see which is the theory (or class
of theories) compatible with the actual observations. Such classification
procedure has been described in detail in \cite{el73}.

The spherical antenna is thus seen to have the {\it capability of furnishing
the analyst sufficient information to discern amongst different competing
theories of GW physics, whenever the wave incidence direction is known
prior to detection}.

\item[{\sf ii)}] The source position is {\it not\/} known at detection
time. This makes things more complex, since the above rotation between the
laboratory and GW frames cannot be performed.

In order to deconvolve the incidence direction in this case, a specific
theory of the GWs {\it must\/} be assumed ---a given choice being made
on the basis of whatever prior information is available or, simply,
dictated by the the decision to probe a particular theory. Wagoner and
Paik \cite{wp77} propose a method which is useful both for GR and BD
theory, their idea being simple and elegant at the same time: since
neither of these theories predicts the excitation of the $m\/$=$\pm$1
quadrupole modes {\it of the wave}, the source position is determined
precisely by the rotation angles which, when applied to the laboratory
axes, cause the amplitudes of those {\it antenna\/} modes to vanish;
the rotated frame is thereby associated to the GW natural frame.

A generalisation of this idea can conceivably be found on the basis of a
detailed ---and possibly rather casuistic--- analysis of the canonical
forms of of the Riemann tensor for a list of theories of gravity, along
the following line of argument: any one particular theory will be
characterised by certain (homogeneous) canonical relationships amongst
the monopole and quadrupole components of the Riemann tensor,
$g^{(S,m)}(t)$, and so enforcement of those relations upon rotation of
the laboratory frame axes should enable determination of the rotation
angles or, equivalently, of the incoming radiation incidence direction.
Scalar-tensor theories e.g. have $g^{(\pm 1)}(t)=0$ in their canonical
forms, hence Wagoner and Paik's proposal for this particular case.

Before any deconvolution procedure is triggered off, however, it is very
important to make sure that it will be {\it viable}. More precisely,
since the transformation from the laboratory to the ultimate canonical
frame is going to be linear, {\it invariants\/} must be preserved. This
means that, even if the source position is unknown, certain theories will
forthrightly be {\it vetoed\/} by the observed $R_{0i0j}(t)$ if their
predicted invariants are incompatible with the observed ones. To give but
an easy example, if $R_{0i0j}(t)$ is {\it observed\/} to have a non-null
trace $R_{0i0i}(t)$, then a veto on GR will be readily served, and
therefore no algorithm based on that theory should be applied.

I would like to make a final remark here. Assume a direction deconvolution
procedure has been successfully carried through to the end on the basis
of certain GW theory, so that the analyst comes up with a pair of numbers
$(\theta,\varphi)$ expressing the source's coordinates in the sky. Of
course, these numbers will represent the {\it actual\/} source position
{\it only if the assumed theory is correct}. Now, how do we know it
{\it is\/} correct? Strictly speaking, ``correctness'' of a scientific
theory is an {\it asymptotic\/} concept ---in the sense that the
possibility always remains open that new facts be eventually discovered
which contradict the theory---, and so {\it reliability\/} of the
estimate $(\theta,\varphi)$ of the source position can only be assessed
in practice in terms of the {\it consistency\/} between the assumed
theory and whatever experimental evidence is available {\it to date},
including, indeed, GW measurements themselves. It is thus very important
to have a method to verify that the estimate $(\theta,\varphi)$ does
not contradict the theory which enabled its very determination.

Such verification is a {\it logical\/} absurdity if only {\it one\/}
measurement of position is available; this happens for instance if the
recorded signal is a {\it short burst\/} of radiation, and so {\it two
antennas\/} are at least necessary to check consistency in that case.
The test would proceed as a check that the time delay between reception
of the signal at both detectors is consistent with the calculated
$(\theta,\varphi)$\footnote{
Note that the two detectors will agree on the same $(\theta,\varphi)$,
even if the assumed theory is wrong, since the sphere deformations will
be the same if caused by the same signal.},
given their relative position and the wave propagation speed predicted by
the assumed theory. If, on the other hand, the signal being tracked is
a {\it long duration\/} signal, then a single antenna may be sufficient
to peform the test by looking at the observed Doppler patterns and
checking them against those expected with the given $(\theta,\varphi)$.

\end{enumerate}

The above considerations have been made ignoring noise in the detector
and monitor systems. A fundamental constraint introduced by noise is
that it makes the antenna {\it bandwidth limited\/} in sensitivity. As
a consequence, any deconvolution procedure is deemed to be incomplete
or, rather, {\it ambiguous\/} \cite{als}, since information about the
signal can possibly be retrieved only within a reduced bandwidth,
whilst the rest will be lost. I thus come to a detailed discussion of
the sensitivity of the spherical GW antenna in the next section.

\section{The sensitivity parameters}

I will consider successively {\it amplitude\/} and {\it energy\/}
sensitivities; the first leads to the concept of {\it transfer
function}, while the second to that of absorption {\it cross
section}. I devote separate subsections to analyse each of them
in some detail.

\subsection{The transfer function}

A widely used and useful concept in linear system theory is that of
{\it transfer function\/} \cite{he68}. It is defined as the Fourier
transform of the system's impulse response, or as the system's
impedance/admittance, and can be inferred from the frequency response
function (\ref{3.18}).

We recall from the previous section that the sphere is a multimode
device ---due to its monopole and five-fold degenerate quadrupole modes.
It appears expedient to define a {\it multimode\/} or {\it vector
transfer function\/} as a useful construct which encompasses all six
different modes into a single conceptual block, according to

\begin{equation}
   {\bf U}({\bf x},\omega) = \sum_\alpha\,{\bf Z}^{(\alpha)}({\bf x},\omega)
               \,G^{(\alpha)}(\omega)          \label{4.1}
\end{equation}

where $G^{(\alpha)}(\omega)$ are the six driving terms $G^{(S,m)}(\omega)$
given in (\ref{3.19}). The transfer function is
${\bf Z}^{(\alpha)}({\bf x},\omega)$, and its ``vector'' character alluded
above is reflected by the {\it multimode index\/} $\alpha$. Looking at
(\ref{3.18}) it is readily seen that

\begin{mathletters}
\label{4.2}
\begin{eqnarray}
  {\bf Z}^{(S)}({\bf x},\omega) & = & \frac{\pi}{i}\,\sum_{n=1}^\infty\,
  \frac{a_n}{\omega_{n0}}\,{\bf u}_{n00}({\bf x})\,\left[
  \delta(\omega\!-\!\omega_{n0}) - \delta(\omega\!+\!\omega_{n0})
  \right]       \label{4.2a}    \\[1.3 em]
  {\bf Z}^{(m)}({\bf x},\omega) & = & \frac{\pi}{i}\,\sum_{n=1}^\infty\,
  \frac{b_n}{\omega_{n2}}\,{\bf u}_{n2m}({\bf x})\,\left[
  \delta(\omega\!-\!\omega_{n2}) - \delta(\omega\!+\!\omega_{n2})
  \right]  \qquad (m=-2,\ldots,2)     \label{4.2b}
\end{eqnarray}
\end{mathletters}

As we observe in these formulae, the sphere's sensitivity to monopole
excitations is governed by $a_n/\omega_{n0}$, and to quadrupole ones by
$b_n/\omega_{n2}$. Closed expressions happen to exist for $a_n\/$ and
$b_n$; using the notation of Appendix B, they are

\begin{mathletters}
\label{4.4}
\begin{eqnarray}
   \frac{a_n}{R} & = & \frac{3\,C(n,0)}{8\pi}\;\frac{j_2(q_{n0}R)}{q_{n0}R}
   \label{4.4a}  \\[1 em]
   \frac{b_n}{R} & = & -\frac{3\,C(n,2)}{8\pi}\,\left[\beta_3(k_{n2}R)\,
   \frac{j_2(q_{n2}R)}{q_{n2}R} - 3\,\frac{q_{n2}}{k_{n2}}\,
   \beta_1(q_{n2}R)\,\frac{j_2(k_{n2}R)}{k_{n2}R}
   \right]     \label{4.4b}
\end{eqnarray}
\end{mathletters}

Numerical investigation of the behaviour of these coefficients shows that
they decay asymptotically as $n^{-2}$:

\begin{equation}
    a_n, b_n \stackrel{n\rightarrow\infty}{\longrightarrow}
    {\rm const}\times n^{-2}          \label{4.6}
\end{equation}

Likewise, it is found that the frequencies $\omega_{n0}\/$ and
$\omega_{n2}\/$ diverge like $n\/$ for large $n$, so that
${\bf Z}^{(\alpha)}({\bf x},\omega)$ drops as $\omega^{-3}$ for
large $\omega$. Figures 6 and 7 display a symbolic plot of
$\omega^3\,{\bf Z}^{(S)}({\bf x},\omega)$ and
$\omega^3\,{\bf Z}^{(m)}({\bf x},\omega)$, respectively, which illustrates
the situation: monopole modes soon reach the asymptotic regime, while
there appear to be 3 subfamilies of quadrupole modes regularly intertwined;
the asymptotic regime for these subfamilies is more irregularly reached.
Note also the perfectly regular alternate changes of phase (by $\pi\/$
radians) in both monopole and each quadrupole family.

The sharp fall in sensitivity of a sphere for higher frequency modes
($n^{-3}$) indicates that only the lowest ones stand a chance of being
obervable in an actual GW antenna. I report in Table I the numerical
values of the relevant parameters for the first few monopole and
quadrupole modes. The reason for the last (fourth) columns will become
clear later.

\begin{table}
\caption{First few monopole (left) and quadrupole (right) sphere
parameters, for a $\sigma\/$=\,0.33 material. First and second columns
on either side of the central line number the modes and give the
corresponding eigenvalue; rows are intertwined in order of ascending
frequency, which is proportional to $kR\/$ ---see (\protect\ref{B.4})
below. Third columns contain the $a_n\/$ and $b_n\/$ coefficients
defined in equations (\protect\ref{3.12a}) and (\protect\ref{3.12b}),
respectively; the fourth columns display the cross section {\it ratios\/}
$(k_{10}a_1/k_{n0}a_n)^2$ and $(k_{12}b_1/k_{n2}b_n)^2$ for higher
frequency modes, respectively, taking as reference the lowest in each
familiy ---cf. equations (\protect\ref{4.23}).}
\begin{tabular}{dddd|dddd}
$n\ \ $ & $k_{n0}R$ & $a_n/R$ & $\sigma_{10}/\sigma_{n0}\ $ &
$\ n\ \ $ & $k_{n2}R$ & $b_n/R$ & $\sigma_{12}/\sigma_{n2}$
\\ \hline \\
  & &  &  & \ 1\ \  & 2.650 &  0.328 & 1                         \\
  & &  &  & \ 2\ \  & 5.088 &  0.106 & 2.61    \\
1\ \  & 5.432 & 0.214  & 1\     &       &       &         &     \\
  & &  &  & \ 3\ \  & 8.617  & $-$1.907$\times\! 10^{-2}$ & 27.95 \\
  & &  &  & \ 4\ \  & 10.917 & $-$9.101$\times\! 10^{-3}$ & 76.42 \\
2\ \  & 12.138 & $-$3.772$\times\! 10^{-2}$ & 6.46\  & & & &    \\
  & &  &  & \ 5\ \  & 12.280 &  1.387$\times\! 10^{-2}$ & 25.99   \\
  & &  &  & \ 6\ \  & 15.347 &  6.879$\times\! 10^{-3}$ & 67.87   \\
3\ \  & 18.492 & 1.600$\times\! 10^{-2}$  & 15.49\    & & & &   \\
\end{tabular}
\end{table}

\subsection{The absorption cross section}

Let us calculate now the energy of the oscillating sphere. We
first define the {\it spectral energy density\/} at frequency $\omega$,
which is naturally given by\footnote{
$T\/$ is the integration time ---assumed very large. The peaks in the
$\delta\/$-functions diverge like $T/\pi$.}

\begin{equation}
  W(\omega) = \frac{1}{T}\;\int_{\rm Solid}\frac{1}{2}\,\omega^2\,
        \left|{\bf U}({\bf x},\omega)\right|^2\,\rho\,d^3x  \label{4.7}
\end{equation}

and can be easily evaluated:

\begin{eqnarray}
  W(\omega) & = & \frac{1}{2}\pi M\,\sum_{n=1}^\infty\,
  a_n^2\,\left|G^{(S)}(\omega)\right|^2
  \left[\delta(\omega\!-\!\omega_{n0}) + \delta(\omega\!+\!
  \omega_{n0})\right]
  + \nonumber \\*[1 em]
  & + & \!\frac{1}{2}\pi M\,\sum_{n=1}^\infty\,
  b_n^2\,\left[\sum_{m=-2}^2\,\left|G^{(m)}(\omega)\right|^2\right]
  \left[\delta(\omega\!-\!\omega_{n2}) + \delta(\omega\!+\!
  \omega_{n2})\right]
  \label{4.8}
\end{eqnarray}

The {\it energy\/} at any one spectral frequency $\omega_{nl}\/$ is
obtained by {\it integration\/} of the spectral density in a narrow
interval around $\omega = \pm\omega_{nl}\/$:

\begin{equation}
  E(\omega_{nl})=\int_{-\omega_{nl}-\varepsilon}^{-\omega_{nl}+\varepsilon}
               + \int_{\omega_{nl}-\varepsilon}^{\omega_{nl}+\varepsilon}
                 \;W(\omega)\,\frac{d\omega}{2\pi}      \label{4.9}
\end{equation}

In particular,

\begin{mathletters}
\label{4.10}
\begin{eqnarray}
  E(\omega_{n0}) & = & \frac{1}{2}\,M\,a_n^2\,\left|G^{(S)}(\omega_{n0})
    \right|^2       \label{4.10a}     \\*[0.5 em]
  E(\omega_{n2}) & = & \frac{1}{2}\,M\,b_n^2\,
  \sum_{m=-2}^2\,\left|G^{(m)}(\omega_{n2})\right|^2    \label{4.10b}
\end{eqnarray}
\end{mathletters}

The sensitivity parameter associated with the vibrational energy of the
modes is the detector's {\it absorption cross section\/}, defined as the
energy it absorbs per unit incident GW spectral flux density, or

\begin{equation}
   \sigma_{\rm abs}(\omega) = \frac{E(\omega)}{\Phi(\omega)}  \label{4.12}
\end{equation}

where $\Phi(\omega)$ is the number of joules per square metre and Hz
carried by the GW at frequency $\omega\/$ as it passes by the antenna.
Thus, for the frequencies of interest,

\begin{mathletters}
\label{4.13}
\begin{eqnarray}
  \sigma_{\rm abs}(\omega_{n0}) & = &
  \frac{1}{2}Ma_n^2\,\frac{\left|G^{(S)}(\omega_{n0})\right|^2}
  {\Phi(\omega_{n0})}  \label{4.13a}    \\*[0.7 em]
  \sigma_{\rm abs}(\omega_{n2}) & = &
  \frac{1}{2}Mb_n^2\,\frac{\sum_{m=-2}^2\,\left|G^{(m)}(\omega_{n2})
  \right|^2}{\Phi(\omega_{n2})}  \label{4.13b}
\end{eqnarray}
\end{mathletters}

These quantities have very precise values, but such values can only be
calculated on the basis of a {\it specific underlying theory of the GW
physics}. In the absence of such theory, neither $\Phi(\omega)$ nor
$G^{(S,m)}(\omega)$ can possibly be calculated, since they are {\it not\/}
theory independent quantities. To date, only GR calculations have been
reported in the literature \cite{wp77,MZ94,clo}. As I will now show,
even though the fractions in the rhs of (\ref{4.13}) are {\it not\/}
theory independent, some very general results can still be obtained
about the sphere's cross section within the context of metric theories
of the gravitational interaction. To do so, it will be necessary to go
into a short digression on the general nature of weak metric GWs.

No matter which is the (metric) theory which happens to be the ``correct
one'' to describe gravitation, it is beyond reasonable doubt that any
GWs reaching the Earth ought to be {\it very weak}. The linear approximation
should therefore be an extremely good one to describe the propagating field
variables in the neighbourhood of the detector. In such circumstances, the
field equations can be derived from a Poincar\'e invariant variational
principle based on an action integral of the type

\begin{equation}
   \int\,{\cal L}(\psi_A,\psi_{A,\mu})\,d^4x      \label{4.15}
\end{equation}

where the Lagrangian density ${\cal L}\/$ is a {\it quadratic\/} functional
of the field variables $\psi_A(x)$ and their space-time derivatives
$\psi_{A,\mu}(x)$; these variables include the metric perturbations
$h_{\mu\nu}\/$, plus any other fields required by the specific theory
under consideration ---e.g. a scalar field in the theory of Brans--Dicke,
etc. The requirement that ${\cal L}\/$ be quadratic ensures that the
Euler--Lagrange equations of motion are {\it linear}.

The energy and momentum transported by the waves can be calculated in
this formalism in terms of the components $\tau^{\mu\nu}$ of the canonical
energy-momentum tensor\footnote{
This tensor is {\it not\/} symmetric in general, but can be
symmetrized by a standard method due to Belinfante \cite{ll85,Barut}. For
the considerations which follow in this paper it is unnecessary to go
into those details, and the {\it canonical\/} form (\ref{4.16}) will be
sufficient.}

\begin{equation}
   \tau^{\mu\nu}({\bf x},t) = \sum_A\,\frac{\partial{\cal L}}
    {\partial\psi_{A,\mu}}\,\psi_A^{,\nu} - {\cal L}\,\eta^{\mu\nu}
    \label{4.16}
\end{equation}

The flux energy density, or Poynting, vector is given by
$S_i = c^2\,\tau^{0i}$, i.e.,

\begin{equation}
   {\bf S}({\bf x},t) = c^3\,\sum_A\,
   \frac{\partial{\cal L}}{\partial\dot\psi_A}\,\nabla\psi_A
   \label{4.17}
\end{equation}

where \ ${\bf\dot{}}\equiv\partial/\partial t$. Any GW hitting the antenna
will be seen plane, due to the enormous distance to the source. If {\bf k}
is the incidence direction (normal to the wave front), then the fields
will depend on the variable $ct\,$$-${\bf k$\cdot$x}, so that the GW
energy reaching the detector per unit time and area is

\begin{equation}
  \phi(t)\equiv {\bf k\!\cdot\!S}({\bf x},t) = -c^2\,\sum_A\,
   \frac{\partial{\cal L}}{\partial\dot\psi_A}\,\dot\psi_A
   \label{4.18}
\end{equation}

where {\bf x} is the sphere's centre position relative to the source
---which is {\it fixed\/}, and so its dependence can be safely dropped
in the lhs of the above expression. The important thing to note in equation
(\ref{4.18}) is that it tells us that $\phi(t)$ {\it can be written as a
quadratic form in the time derivatives of the fields $\psi_A$}. As a
consequence, the spectral density $\Phi(\omega)$, defined by

\begin{equation}
   \int_{-\infty}^\infty\,\phi(t)\,dt = \int_0^\infty\,
   \Phi(\omega)\,\frac{d\omega}{2\pi}      \label{4.20}
\end{equation}

can be ascertained to factorise as

\begin{equation}
   \Phi(\omega) = \omega^2\,\Phi_0(\omega)    \label{4.21}
\end{equation}

where $\Phi_0(\omega)$ is again a {\it quadratic\/} function of the
Fourier transforms $\Psi_A(\omega)$ of the fields $\psi_A$. On the
other hand, the functions $G^{(S,m)}(\omega)$ in (\ref{4.13}) which,
it is recalled, are the Fourier transforms of $g^{(S,m)}(t)$ in
(\ref{3.8}), contain {\it second\/} order derivatives of the
{\it metric\/} fields $h_{\mu\nu}$, and therefore of {\it all\/} the
fields $\psi_A$ as a result of the theory's field equations. Since we
are considering {\it plane wave\/} solutions to those equations, all
derivatives can be reduced to {\it time\/} derivatives ---just like
in (\ref{4.18}) above. We can thus write

\begin{equation}
   G^{(S,m)}(\omega) = -\omega^2\,\Psi^{(S,m)}(\omega)   \label{4.22}
\end{equation}

with $\Psi^{(S,m)}(\omega)$ suitable {\it linear\/} combinations of the
$\Psi_A(\omega)$. Replacing the last two equations into (\ref{4.13})
and manipulating dimensions expediently, we come to the remarkable
result that

\begin{mathletters}
\label{4.23}
\begin{eqnarray}
  \sigma_{\rm abs}(\omega_{n0}) & = &
  K_S(\aleph)\,\frac{GMv_t^2}{c^3}\,(k_{n0}a_n)^2
  \label{4.23a}     \\*[0.5 em]
  \sigma_{\rm abs}(\omega_{n2}) & = &
  K_Q(\aleph)\,\frac{GMv_t^2}{c^3}\,(k_{n2}b_n)^2  \label{4.23b}
\end{eqnarray}
\end{mathletters}

where $v_t^2\,$$\equiv\,$(2+2$\sigma)^{-1}$$\,v_s^2$, $v_s\/$ being the
speed of sound in the detector's material, and $\sigma\/$ its Poisson
ratio; $G\/$ is the Gravitational constant. The ``remarkable'' about
the above is that the coefficients $K_S(\aleph)$ and $K_Q(\aleph)$
{\it are independent of frequency\/}: they {\it exclusively depend on the
underlying gravitation theory}, which I symbolically denote by $\aleph$.
To see that this is the case, it is enough to consider a monochromatic
incident wave: since the coefficients $K_S(\aleph)$ and $K_Q(\aleph)$
happen to be {\it invariant\/} with respect to field amplitude
{\it scalings}, this means they will {\it only\/} depend on the amplitudes'
relative weights, i.e., on the field equations' {\it specific structure}.

By way of example, it is interesting to see what the results for General
Relativity (GR) and Brans--Dicke (BD) theory are. After somehow lengthy
algebra it is found that

\begin{equation}
   \aleph = {\rm GR} \Rightarrow \left\{\begin{array}{l}
          K_S(\aleph) = 0 \\[0.7 em]
          K_Q(\aleph) = \mbox{\large $\frac{16\,\pi^2}{15}$}
          \end{array}\right.   \label{4.25}
\end{equation}

and

\begin{equation}
   \aleph = {\rm BD} \Rightarrow \left\{\begin{array}{l}
          K_S(\aleph) = \mbox{\large $\frac{8\,\pi^2}{9}$}\,
          (3+2\Omega)^{-2}\,k\,\left[1+\mbox{\large $\frac
          {k\Omega}{(3+2\Omega)^2}$}\right]^{-1}  \\*[1.3 em]
          K_Q(\aleph) = \mbox{\large $\frac{16\,\pi^2}{15}$}\,\left[1 +
          \frac{1}{6}\,(3+2\Omega)^{-2}\,k\right]
          \left[1+\mbox{\large $\frac{k\Omega}{(3+2\Omega)^2}$}\right]^{-1}
          \end{array}\right.   \label{4.26}
\end{equation}

In the latter formulae, $\Omega$ is the usual Brans--Dicke parameter
$\omega\/$ \cite{bd61}, renamed here to avoid confusion with
{\it frequency}, and $k\/$ is a dimensionless parameter, generally
of order one, depending on the source's properties \cite{lobo}. As
is well known, GR is obtained in the limit $\Omega\rightarrow\infty$
of BD \cite{we72}; the quoted results are of course in agreement with
that limit.

Incidentally, an interesting consequence of the above equations is
this: though not explicitly shown in this paper (see, however, reference
\cite{wp77}), the presence of a scalar field in the theory of Brans and
Dicke causes {\it not only the monopole\/} sphere's modes to be excited,
{\it but also the {\rm $m\/$=0} quadrupole ones}; what we see in equations
(\ref{4.26}) is that {\it precisely\/} 5/6 of the total energy extracted
from the scalar wave goes into the antenna's monopole modes, whilst there
is still a remaining 1/6 which is communicated to the quadrupoles,
independently of the values of $\Omega$ and $k$\footnote{
Note however that since monopole and quadrupole detector modes occur at
different frequencies, this particular {\it distribution\/} of energy
may not be seen if the sphere's vibrations are monitored at a single
resonance.}.
This somehow non-intuitive result finds its explanation in the structure
of the Riemann tensor in BD theory, in which the {\it excess\/}
$R_{0i0j}\/$ with respect to General Relativity happens {\it not\/} to be
proportional to the scalar part $E_{ij}^{(S)}$, but to a combination of
$E_{ij}^{(S)}$ \mbox{and $E_{ij}^{(0)}$.}

Equations (\ref{4.23}) show that, no matter which is the gravity theory
assumed, the sphere's absorption cross sections for higher modes
{\it scale\/} as the successive coefficients $(k_{n0}a_n)^2$ and
$(k_{n2}b_n)^2$ for monopole and quadrupole modes, respectively. In
particular, the result quoted in \cite{clo} that cross section for the
second quadrupole mode is 2.61 times less than that for the first,
assuming GR, is in fact valid, as we now see, {\it independently of
which is the\/ {\rm (metric)} theory of gravity actually governing GW
physics}. The fourth columns in Table I display these scaling properties.
It is seen that the drop in cross section from the first to the second
monopole mode is as high as 6.46. It should however be stressed that the
frequency of such mode would be over 4 kHz for a (likely) sphere whose
fundamental {\it quadrupole\/} frequency be 900 Hz~\cite{clo}. Note
finally the asymptotic cross section drop as $n^{-2}\/$ for large
$n\/$ ---cf. equation (\ref{4.6}) and the ensuing paragraph.

\section{Conclusion}

The main purpose of this paper has been to set up a {\it sound\/}
mathematical formalism to tackle with as much generality as possible any
questions related to the interaction between a resonant antenna and a weak
incoming GW, with much special emphasis on the homogeneous {\it sphere}.
New results have been found along this line, such as the {\it scaling\/}
properties of cross sections for higher frequency modes, or the sensitivity
of the antenna to arbitrary metric GWs; also, new ideas have been put
forward regarding the {\it direction deconvolution\/} problem within the
context of an arbitrary metric theory of GW physics. Less spectacularly,
the full machinery has also been applied to produce {\it independent\/}
checks of previously published results.

The whole investigation reported herein has been developed with no
{\it a priori\/} assumptions about any specific (metric) theory of the
GWs, and is therefore {\it very\/} general. ``Too general solutions'' are
often impractical in science; here, however, the ``very general'' appears
to be rather ``cheap'', as seen in the results expressed by the equations
of section 3 above. An immediate consequence is that solid elastic detectors
of GWs (and, in particular, spheres) offer, as a matter of principle, the
possibility of probing {\it any\/} given theory of GW physics with just as
much effort as it would take, e.g., to probe General Relativity: the vector
transfer function of section 4 supplies the requisite theoretical vehicle
for the purpose.

An important question, however, has not been considered in this paper.
This is the {\it transducer\/} problem: the sphere's oscillations can
only be revelaed to the observer by means of suitable (usually
electromechanical) transducers. These devices, however, are not
{\it neutral\/}, i.e., they {\it couple\/} to the antenna's motions,
thereby excercising a back action on it which must be taken into
consideration if one is to correctly interpret the system's readout.
Preliminary studies and proposals have already been published \cite{jm93},
but further work is clearly needed for a more thorough understanding of
the problems involved.

Progress in this direction is currently being made ---which I expect to
report on shortly. The formalism developed in this paper provides basic
support to that further work.

\acknowledgements{
It is a pleasure for me to thank Eugenio Coccia for his critical reading
and comments on the manuscript, and to JMM Senovilla for having devoted
a part of his time to enlightening discussions with me on the geometrical
nature and properties of metric GWs. I also want to express gratitude to
M Montero and JA Ortega for their assistance during the first satges of
this work. I have received support from the Spanish Ministry of Education
through contract number PB93--1050.}

\appendix

\section{}

Let ${\bf e}_x,{\bf e}_y,{\bf e}_z$ be three orthonormal Cartesian vectors
defining the sphere's laboratory reference frame. We define the equivalent
triad

\begin{equation}
  {\bf e}^{(0)}={\bf e}_z \qquad,\qquad {\bf e}^{(\pm 1)}=
  \mbox{\large $\frac{1}{\sqrt{2}}$}\,({\bf e}_x\pm i{\bf e}_y)
  \label{A.1}
\end{equation}

having the properties

\begin{equation}
   {\bf e}^{*(m')}\cdot{\bf e}^{(m)} = \delta_{m'm}
   \qquad ,\qquad m,m'=-1,0,1   \label{A.2}
\end{equation}

We say that the vectors (\ref{A.1}) are the {\it natural basis\/} for the
$l\/$=1 irreducible representation of the rotation group; they behave
under arbitrary rotations precisely like the spherical harmonics
$Y_{1m}(\theta,\varphi)$. In particular, if a rotation of angle
$\alpha\/$ around the \mbox{$z\/$-axis} is applied to the original frame,
then

\begin{equation}
  {\bf e}^{(\pm 1)}\rightarrow\exp(\pm i\alpha)\,{\bf e}^{(\pm 1)}
  \qquad,\qquad {\bf e}^{(0)}\rightarrow{\bf e}^{(0)}   \label{A.3}
\end{equation}

Higher rank tensors have specific multipole characteristics depending on
the number of tensor indices, and the above basis lends itself to reveal
those characteristics, too. For example, the five dimensional linear space
of traceless symmetric tensors supports the $l\/$=2 irreducible
representation of the rotation group, while a tensor's trace is an
invariant. A general symmetric tensor can be expressed as an ``orthogonal''
sum of a traceless symmetric tensor and a multiple of the unit tensor.
A convenient basis to expand any such tensor is the following:

\begin{mathletters}
\label{A.4}
\begin{eqnarray}
  & {\bf e}^{(1)}\otimes{\bf e}^{(1)}\qquad,\qquad
    {\bf e}^{(-1)}\otimes{\bf e}^{(-1)} & \label{A.4a} \\*[0.5 em]
  & {\bf e}^{(0)}\otimes{\bf e}^{(1)}+{\bf e}^{(1)}\otimes{\bf e}^{(0)}
                               \ \ ,\ \ 
    {\bf e}^{(0)}\otimes{\bf e}^{(-1)}+{\bf e}^{(-1)}\otimes{\bf e}^{(0)} & 
                             \label{A.4b} \\*[0.5 em]
  & {\bf e}^{(1)}\otimes{\bf e}^{(-1)}+{\bf e}^{(-1)}\otimes{\bf e}^{(1)} -
    2\,{\bf e}^{(0)}\otimes{\bf e}^{(0)} & \label{A.4c} \\*[0.5 em]
  & {\bf e}^{(1)}\otimes{\bf e}^{(-1)}+{\bf e}^{(-1)}\otimes{\bf e}^{(1)} +
    {\bf e}^{(0)}\otimes{\bf e}^{(0)} & \label{A.4d}
\end{eqnarray}
\end{mathletters}

The elements (\ref{A.4a}) get multiplied by e$^{\pm 2i\alpha}\/$ in a
rotation of angle $\alpha\/$ around the \mbox{$z\/$-axis}, respectively,
the (\ref{A.4b}) by e$^{\pm i\alpha}\/$, and (\ref{A.4c}) and
(\ref{A.4d}) are invariant, as is readily seen. These properties define
the ``spin characteristics'' of the corresponding tensors. Also, the
five elements (\ref{A.4a})--(\ref{A.4c}) are {\it traceless\/} tensors,
while (\ref{A.4d}) is the {\it unit\/} tensor. Any symmetric tensor can
be expressed as a linear combination of the six (\ref{A.4}), and the
respective coefficients carry the information about the weights of the
different monopole and quadrupole components of the tensor.

Equations (\ref{3.4}) in the text are the matrix representation of the
above tensors in the Cartesian basis ${\bf e}_x,{\bf e}_y,{\bf e}_z$,
except that they are multiplied by suitable coefficients to ensure that
the conditions

\begin{equation}
  E_{ij}^{(S)}\,n_i n_j = Y_{00}(\theta,\varphi)\ \ ,\ \ 
  E_{ij}^{(m)}\,n_i n_j = Y_{2m}(\theta,\varphi)  \label{A.8}
\end{equation}

where {\bf n}$\,\equiv\,${\bf x}/$|{\bf x}|$ is the radial unit vector,
hold. They are arbitrary, but expedient for the calculations in this
paper. The following {\it orthogonality\/} relations can be easily
established:

\begin{equation}
  E_{ij}^{*(m')}\,E_{ij}^{(m)} = 
  \mbox{\large $\frac{15}{8\pi}$}\,\delta_{m'm}\ \ ,\ \ 
  E_{ij}^{(S)}\,E_{ij}^{(m)} = 0\ \ ,\ \ 
  E_{ij}^{(S)}\,E_{ij}^{(S)} = \mbox{\large $\frac{3}{4\pi}$}
  \label{A.9}
\end{equation}

with the indices $m$,$m'$ running from $-2$ to 2, and with an understood
sum over the repeated $i\/$ and $j$. It is also easy to prove the
{\it closure\/} properties

\begin{equation}
  E_{ij}^{(S)}\,E_{kl}^{(S)}\ +\ \frac{2}{5}\,
  \sum_{m=-2}^2\,E_{ij}^{*(m)}\,E_{kl}^{(m)} = 
  \mbox{\large $\frac{3}{8\pi}$}\,(\delta_{ik}\delta_{jl}+
  \delta_{il}\delta_{jk})   \label{A.10}
\end{equation}

Equations (\ref{A.9}) and (\ref{A.10}) constitute the {\it completeness\/}
equations of the \mbox{$E\/$-matrix} basis of Euclidean symmetric tensors.

\section{}

This Appendix is intended to give a rather complete summary of the
frequency spectrum and eigenmodes of a uniform elastic sphere. Although
this is a classical problem in Elasticity Theory~\cite{lo44}, some of
the results which follow have never been published so far. Also, its
scope is to serve as reference for notation, etc., in future work.

The uniform\footnote{
By {\it uniform\/} I mean its density $\varrho\/$ is constant
throughout the solid in the unperturbed state.}
elastic sphere's normal modes are obtained as the solutions to the
eigenvalue equation 

\begin{equation}
   \mu\nabla^2{\bf u} + (\lambda+\mu)\,\nabla(\nabla{\bf \cdot}{\bf u})
      = - \omega^2\varrho{\bf u}     \label{B.0}
\end{equation}
with the boundary conditions that its surface be free of any tensions
and/or tractions; this is expressed by the equations~\cite{ll70}

\begin{equation}
   \sigma_{ij}\,n_j = 0 \qquad {\rm at}\ \ r\!=\!R     \label{B.1}
\end{equation}
where $R\/$ is the sphere's radius, {\bf n} the outward normal, and
$\sigma_{ij}$ the {\it stress\/} tensor

\begin{equation}
   \sigma_{ij} = \lambda\,u_{kk}\delta_{ij} + 2\mu\,u_{ij}   \label{B.1a}
\end{equation}
with $u_{ij}\equiv{\mbox{\normalsize $\frac{1}{2}$}}(u_{i,j}+u_{j,i})$,
the {\it strain\/} tensor, and $\lambda,\mu\/$ the Lam\'e
coefficients~\cite{ll70}.

Like any differentiable vector field, {\bf u}({\bf x}) can be expressed as
a sum of an irrotational vector and a divergence-free vector,

\begin{equation}
  {\bf u}({\bf x}) = {\bf u}_{\rm irrot.}({\bf x}) +
  {\bf u}_{\rm div-free}({\bf x})\ ,
  \label{B.1b}
\end{equation}
say; on substituting this into equation~(\ref{B.0}), and after a few
easy manipulations, one can see that

\begin{equation}
  (\nabla^2 + k^2)\,{\bf u}_{\rm div-free}({\bf x}) = 0\ ,\qquad
  (\nabla^2 + q^2)\,{\bf u}_{\rm irrot.}({\bf x}) = 0
  \label{B.1c}
\end{equation}
where

\begin{equation}
   k^2\equiv\frac{\varrho\omega^2}{\mu} \ ,\qquad
   q^2\equiv\frac{\varrho\omega^2}{\lambda+2\mu}
   \label{B.4}
\end{equation}

Now the irrotational component can generically be expressed as
the {\em gradient\/} of a scalar function, i.e.,

\begin{equation}
   {\bf u}_{\rm irrot.}({\bf x}) = \nabla\phi({\bf x})
  \label{B.1d}
\end{equation}
while there are {\em two\/} linearly independent divergence-free components
which, as can be readily verified, are

\begin{equation}
  {\bf u}_{\rm div-free}^{(1)}({\bf x}) = {\bf L}\psi^{(1)}({\bf x})
  \ ,\ \ \ \mbox{and}\ \ \ \ 
  {\bf u}_{\rm div-free}^{(2)}({\bf x}) =
          \nabla\!\times\!{\bf L}\psi^{(2)}({\bf x})
  \label{B.1e}
\end{equation}
where ${\bf L}\equiv-i{\bf x}\!\times\!\nabla$ is the ``angular momentum''
operator, cf.~\cite{Ed60}, and $\psi^{(1)}$ and $\psi^{(2)}$ are also
scalar functions. If~(\ref{B.1d}) and~(\ref{B.1e}) are now respectively
substituted in~(\ref{B.1c}), it is found that $\phi({\bf x})$,
$\psi^{(1)}({\bf x})$, and $\psi^{(2)}({\bf x})$ satisfy Helmholtz equations:

\begin{equation}
  (\nabla^2 + k^2)\,\psi({\bf x}) = 0\ ,\qquad
  (\nabla^2 + q^2)\,\phi({\bf x}) = 0
   \label{B.1f}
\end{equation}
where $\psi({\bf x})$ stands for either $\psi^{(1)}({\bf x})$ or
$\psi^{(2)}({\bf x})$. Therefore

\begin{equation}
  \phi({\bf x})=j_l(qr)\,Y_{lm}({\bf n}) \qquad,\qquad
  \psi({\bf x})=j_l(kr)\,Y_{lm}({\bf n})     \label{B.3}
\end{equation}
in order to ensure regularity at the centre of the sphere, $r\/$=0. Here,
$j_l\/$ is a {\it spherical\/} Bessel function ---see~\cite{ab72} for
general conventions on these functions---, and $Y_{lm}\/$ a spherical
harmonic~\cite{Ed60}. Finally thus,

\begin{equation}
  {\bf u}({\bf x}) = \frac{C_0}{q^2}\,\nabla\phi({\bf x})
		   + \frac{iC_1}{k}\,{\bf L}\psi({\bf x})
		   + \frac{iC_2}{k^2}\,\nabla\!\times\!{\bf L}\psi({\bf x})
  \label{B.2}
\end{equation}
where $C_0,C_1,C_2$ are three constants which will be determined by the
boundary conditions~(\ref{B.1}) (the denominators under them have been
included for notational convenience). After lengthy algebra, those
conditions can be expressed as the following system of linear equations:

\begin{mathletters}
\label{B.5}
\begin{eqnarray}
  \left[\beta_2(qR)-\mbox{\large $\frac{\lambda}{2\mu}$}\,
  q^2R^2\,\beta_0(qR)\right]\,C_0
  - l(l+1)\,\beta_1(kR)\,C_2 & = & 0   \label{B.5a} \\*[0.5 em]
  \beta_1(kR)\,C_1 & = & 0             \label{B.5b} \\*[0.5 em]
  \beta_1(qR)\,C_0 - \left[\mbox{\large $\frac{1}{2}$}\,\beta_2(kR)+
  \left\{\mbox{\large $\frac{l(l+1)}{2}$}-1\right\}\,
  \beta_0(kR)\right]\,C_2 & = & 0
  \label{B.5c}
\end{eqnarray}
\end{mathletters}
where

\begin{equation}
  \beta_0(z)\equiv\frac{j_l(z)}{z^2}\ ,\ \ \
  \beta_1(z)\equiv\frac{d}{dz}\left[\frac{j_l(z)}{z}\right]\ ,\ \ \
  \beta_2(z)\equiv\frac{d^2}{dz^2}\left[j_l(z)\right]  \label{B.8}
\end{equation}

There are clearly {\it two\/} families of solutions to~(\ref{B.5}):

\begin{enumerate}
\item[{\sf i)}] {\it Toroidal\/} modes. These are characterised by

\begin{equation}
   \beta_1(kR)=0\ ,\qquad C_0=C_2=0      \label{B.11}
\end{equation}

The frequencies of these modes are {\it independent\/} of $\lambda$, and
thence independent of the material's Poisson ratio. Their amplitudes are

\begin{equation}
   {\bf u}_{nlm}^T({\bf x})=T_{nl}(r)\,i{\bf L}Y_{lm}({\bf n})
   \label{B.12}
\end{equation}
with

\begin{equation}
   T_{nl}(r)=C_1(n,l)\,j_l(k_{nl}r)      \label{B.13}
\end{equation}
and $C_1(n,l)$ a dimensionless normalisation constant determined by the
general formula~(\ref{2.10}); $k_{nl}R\/$ is the $n\/$-th root of the
first equation~(\ref{B.11}) for a given $l$.

\item[{\sf ii)}] {\it Spheroidal\/} modes. These correspond to

\begin{equation}
   \det\left(\begin{array}{cc}
   \beta_2(qR)-\mbox{\large $\frac{\lambda}{2\mu}$}\,q^2R^2\,\beta_0(qR) &
   l(l+1)\,\beta_1(kR)   \\
   \beta_1(qR) & \mbox{\large $\frac{1}{2}$}\,\beta_2(kR)+
   \left\{\mbox{\large $\frac{l(l+1)}{2}$}-1\right\}\,\beta_0(kR)
   \end{array}\right) = 0      \label{B.14}
\end{equation}
and $C_1\/$\,=\,0. The frequencies of these modes {\it do\/} depend on
the Poisson ratio, and their amplitudes are

\begin{equation}
   {\bf u}_{nlm}^P({\bf x}) = A_{nl}(r)\,Y_{lm}({\bf n})\,{\bf n}
   - B_{nl}(r)\,i{\bf n}\!\times\!{\bf L}Y_{lm}({\bf n}) \label{B.15}
\end{equation}
where $A_{nl}(r)$ and $B_{nl}(r)$ have the somewhat complicated form

\begin{mathletters}
\label{B.16}
\begin{eqnarray}
  A_{nl}(r) & = & C(n,l)\left[\beta_3(k_{nl}R)\,j_l'(q_{nl}r)
  -l(l+1)\,\frac{q_{nl}}{k_{nl}}\,\beta_1(q_{nl}R)\,
  \frac{j_l(k_{nl}r)}{k_{nl}r}\right]  \label{B.16a} \\*[0.8 em]
  B_{nl}(r) & = & C(n,l)\left[\beta_3(k_{nl}R)\,
  \frac{j_l(q_{nl}r)}{q_{nl}r}-\frac{q_{nl}}{k_{nl}}\,\beta_1(q_{nl}R)\,
  \frac{\left\{k_{nl}r\,j_l(k_{nl}r)\right\}'}{k_{nl}r}\right]
  \label{B.16b}
\end{eqnarray}
\end{mathletters}
with accents denoting derivatives with respect to implied (dimensionless)
arguments,

\begin{equation}
   \beta_3(z)\equiv\mbox{\large $\frac{1}{2}$}\,\beta_2(z) + 
   \left\{\mbox{\large $\frac{l(l+1)}{2}$}-1\right\}\,\beta_0(z)
   \label{B.18}
\end{equation}
and $C(n,l)$ a new normalisation constant. It is understood that
$q_{nl}\/$ and $k_{nl}\/$ are obtained after the (transcendental)
equation~(\ref{B.14}) has been solved for $\omega\/$ ---cf.
equation~(\ref{B.4}).
\end{enumerate}

In actual practice equations~(\ref{B.11}) and~(\ref{B.14}) are solved for
the {\em dimensionless\/} quantity $kR$, which will hereafter be called
the {\em eigenvalue\/}. In view of~(\ref{B.4}), the relationship between
the latter and the measurable frequencies (in Hz) is given by

\begin{equation}
   \nu\equiv\frac\omega{2\pi} = \left(\frac\mu{\varrho\,R^2}\right)^{1/2}\,
   \frac{kR}{2\pi}
   \label{B.19}
\end{equation}

It is more useful to express the frequencies in terms of the Poisson ratio,
$\sigma$, and of the speed of sound $v_{\rm s}$ in the selected material.
For this the following formulas are required ---see e.g.~\cite{ll70}:

\begin{equation}
   v_{\rm s} = \sqrt{\frac {\sf Y}\varrho}
   \label{B.20}
\end{equation}
where {\sf Y} is the {\em Young modulus\/}, related to the Lam\'e
coefficients and the Poisson ratio by

\begin{equation}
   {\sf Y} = \frac{(3\lambda + 2\mu)\,\mu}{\lambda + \mu} =
	     2\,(1+\sigma)\,\mu\ ,\qquad
   \sigma\equiv\frac\lambda{2(\lambda + \mu)}
   \label{B.21}
\end{equation}

Hence,

\begin{equation}
   \nu = \frac{(kR)}{2\pi\,\sqrt{1+\sigma}}\,\frac{v_{\rm s}}R
   \label{B.22}
\end{equation}

Equation~(\ref{B.22}) provides a suitable transformation formula from
abstract number eigenvalues $(kR)$ into physical frequencies $\nu$,
for given material's properties and sizes.

Tables~\ref{tab2} and~\ref{tab3} respectively display a set of values
of $(kR)$ for {\em toroidal\/} and {\em spheroidal\/} modes. While GWs
can only couple to quadrupole and monopole modes, it is important to have
some detailed knowledge of analytical results, as the sphere's frequency
spectrum is rather involved. It often happens, both in numerical
simulations and in experimental determinations, that it is very difficult
to disentangle the wealth of observed frequency lines, and to correctly
associate them with the corresponding eigenmode. Complications are
further enhanced by partial degeneracy lifting found in practice (due
to broken symmetries), which result in even more frequency lines in the
spectrum. Accurate analytic results should therefore be very helpful
to assist in frequency identification tasks.

\begin{table}[t]
\caption{List of a few {\it spheroidal eigenvalues\/}, ordered in
columns of ascending harmonics for each multipole value. Spheroidal
eigenvalues depend on the sphere's material Poisson ratio ---although
this dependence is weak. In this table, values are given for
$\sigma=0.33$. Note that the table contains {\it all\/} eigenvalues less
than or equal to 11.024 yet is not exhaustive for values larger than
that one; this would require to stretch the table horizontally beyond
$l\/$\,=\,10  ---see Figure I for a qualitative inspection of trends
in eigenvalue progressions.   \label{tab3}}

\begin{center}
\begin{tabular}{c|ccccccccccc}
$n$ & $l=0$   & $l=1$   & $l=2$   & $l=3$   & $l=4$   & $l=5$ &
      $l=6$   & $l=7$   & $l=8$   & $l=9$   & $l=10$ \\
\hline
 1  & 5.4322  & 3.5895  & 2.6497  & 3.9489  & 5.0662  & 6.1118
    & 7.1223  & 8.1129  & 9.0909  & 10.061  & 11.024 \\
 2  & 12.138  & 7.2306  & 5.0878  & 6.6959  & 8.2994  & 9.8529
    & 11.340  & 12.757  & 14.111  & 15.410  & 16.665 \\
 3  & 18.492  & 8.4906  & 8.6168  & 9.9720  & 11.324  & 12.686
    & 14.066  & 15.462  & 16.867  & 18.272  & 19.664 \\
 4  & 24.785  & 10.728  & 10.917  & 12.900  & 14.467  & 15.879
    & 17.243  & 18.589  & 19.930  & 21.272  & 22.619 \\
 5  & 31.055  & 13.882  & 12.280  & 14.073  & 16.125  & 18.159
    & 19.997  & 21.594  & 23.043  & 24.426  & 25.778
\end{tabular}
\end{center}
\end{table}

\begin{table}
\caption{List of a few {\it toroidal eigenvalues\/}, ordered in columns
of ascending harmonics for each multipole value. Unlike spheroidal
eigenvalues, toroidal eigenvalues are independent of the sphere's material
Poisson ratio. Note that the table contains {\it all\/} eigenvalues less
than or equal to 12.866 yet is not exhaustive for values larger than
that one; this would require to stretch the table horizontally beyond
$l\/$\,=\,11 ---see Figure II for a qualitative inspection of trends
in eigenvalue progressions.   \label{tab2}}

\begin{center}
\begin{tabular}{c|ccccccccccc}
$n$ & $l=1$   & $l=2$   & $l=3$   & $l=4$   & $l=5$   & $l=6$ &
      $l=7$   & $l=8$   & $l=9$   & $l=10$  & $l=11$ \\
\hline
 1  & 5.7635  & 2.5011  & 3.8647  & 5.0946  & 6.2658
    & 7.4026  & 8.599   & 9.6210  & 10.711  & 11.792  & 12.866 \\
 2  & 9.0950  & 7.1360  & 8.4449  & 9.7125  & 10.951
    & 12.166  & 13.365  & 14.548  & 15.720  & 16.882  & 18.035 \\
 3  & 12.323  & 10.515  & 11.882  & 13.211  & 14.511
    & 15.788  & 17.045  & 18.287  & 19.515  & 20.731  & 21.937 \\
 4  & 15.515  & 13.772  & 15.175  & 16.544  & 17.886
    & 19.204  & 20.503  & 21.786  & 23.055  & 24.310  & 25.555 \\
 5  & 18.689  & 16.983  & 18.412  & 19.809  & 21.181
    & 22.530  & 23.860  & 25.174  & 26.473  & 27.760  & 29.035
\end{tabular}
\end{center}
\end{table}

In Figures 1 and 2 a symbolic line diagramme of the two families of
frequencies of the sphere's spectrum is presented. Spheroidal eigenvalues
have been plotted for the Poisson ratio $\sigma\/$=0.33. Although only
the $l\/$=0 and $l\/$=2 {\it spheroidal\/} series couple to GW tidal
forces, the plots include other eigenvalues, as they can be useful
both in bench experiments ---cf. equation~(\ref{2.19}) above---
and for vetoing purposes in a spherical antenna.

Figures 3, 4 and 5 contain plots of the first three monopole and
quadrupole functions $T_{nl}(r)$, $A_{nl}(r)$ and $B_{nl}(r)$, always for
$\sigma\/$=0.33. $T_{n0}(r)$ and $B_{n0}(r)$ have however been omitted;
this is because they are multiplied by an identically zero angular
coefficient in the amplitude formulae~(\ref{B.12}) and~(\ref{B.15}).
Indeed, monopole vibrations are spherically symmetric, i.e., purely radial.


\newpage
\begin{center}
{\large\bf List of Figures}
\end{center}
\vspace{1.1 em}\hspace{1.7 em}

{\bf Figure 1\ \ } The homogeneous sphere {\it spheroidal\/} eigenvalues
for a few {\it multipole\/} families. Only the $l\/$=0 and $l\/$=2 families
couple to metric GWs, so the rest are given for completeness and
non-directly-GW uses. Note that there are {\it fewer\/} monopole than any
other $l\/$-pole modes. The lowest frequency is the first {\it quadrupole\/}.
The diagramme corresponds to a sphere with Poisson ratio $\sigma\/$=0.33.
Frequencies can be obtained from the plotted values through equation
(\protect\ref{B.4}) for any specific case.

\vspace{1.2 em}

{\bf Figure 2\ \ } The homogeneous sphere {\it toroidal\/} eigenvalues.
None of these couple to GWs, but knowledge of them can be useful for
{\it vetoing\/} purposes. These eigenvalues are {\it independent\/}
of the material's Poisson ratio. To obtain actual frequencies from
plotted values, use (\ref{B.4}). The lowest {\it toroidal\/}
eigenvalue is \mbox{$kR=2.5011$}, with $l\/$=2, and happens to be the
{\it absolute minimum\/} sphere's eigenvalue. Compared to the
{\it spheroidal\/} \mbox{$kR=2.6497$}, also with $l\/$=2, its
frequency is 5.61\%\ smaller. Note also that there are no monopole
toroidal modes.

\vspace{1.2 em}

{\bf Figure 3\ \ } First three {\it spheroidal monopole\/} radial
functions $A_{n0}(r)$ ($n=1,2,3$), equation (\ref{B.16a}).

\vspace{1.2 em}

{\bf Figure 4\ \ } First three {\it spheroidal quadrupole\/} radial
functions $A_{n2}(r)$ (continuous line) and $B_{n2}(r)$ (broken
line) ($n=1,2,3$), equations (\ref{B.16}).

\vspace{1.2 em}

{\bf Figure 5\ \ } First three {\it toroidal quadrupole\/} radial functions
$T_{n2}(r)$ ($n=1,2,3$), equation (\protect\ref{B.13}). A common feature
to these radial functions (also in the two previous Figures) is that they
present a {\it nodal\/} point at the origin ($r=0$), while the sphere's
surface ($r/R=1$) has a non-zero amplitude value, which is largest (in
absolute value) for the lowest $n\/$ in each group.

\vspace{1.2 em}

{\bf Figure 6\ \ } The {\it scalar\/} component
${\bf Z}^{(S)}({\bf x},\omega)$ of the {\it multimode\/} transfer
function, (\protect\ref{4.2a}). The diagramme actually displays
$\omega^3\,{\bf Z}^{(S)}({\bf x},\omega)$, so asymptotic behaviours
are better appreciated. It is given in units of $\mu/\rho R\/$, and
a factor $(\pi/i)\,{\bf u}_{n00}({\bf x})$, the eigenmode amplitude,
has been omitted, too. $\delta\/$-function amplitudes are symbolically
taken as 1. Note that the asymptotic regime, given by equation
(\ref{4.6}), is quickly reached.

\vspace{1.2 em}

{\bf Figure 7\ \ } The {\it quadrupole\/} component
${\bf Z}^{(m)}({\bf x},\omega)$ of the {\it multimode\/} transfer
function, (\ref{4.2b}). The same prescriptions of Figure~6 apply
here; the plot is therefore {\it independent\/} of the value of $m\/$.
Note the presence of {\it three\/} subfamilies of peaks; asymptotic
regimes are reached with variable speed for these subfamilies, and less
rapidly than for monopole modes, anyway.

\newpage

\begin{figure}
\psfig{file=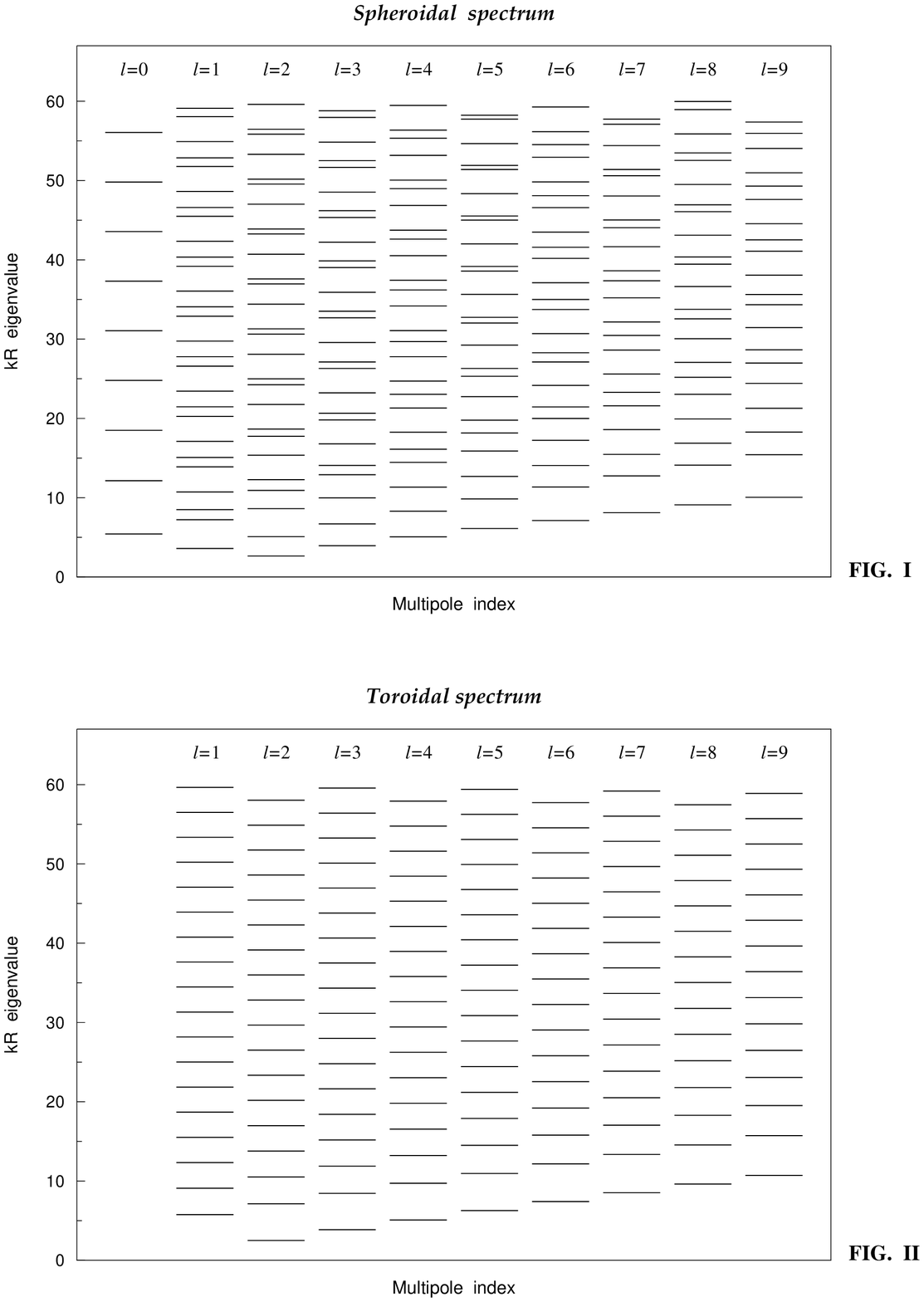,width=16.2cm}
\end{figure}

\begin{figure}
\psfig{file=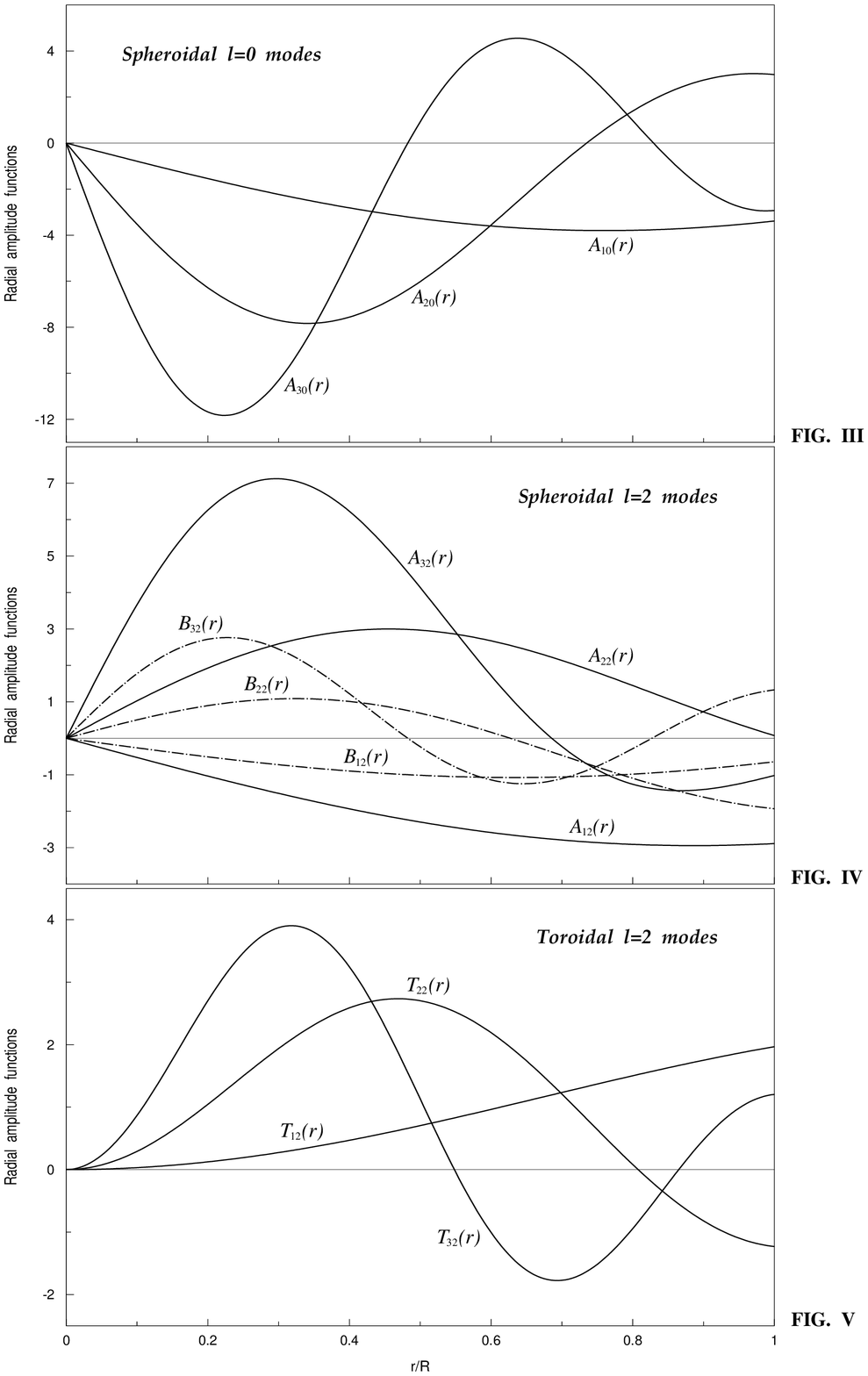,width=16.2cm}
\end{figure}

\begin{figure}
\psfig{file=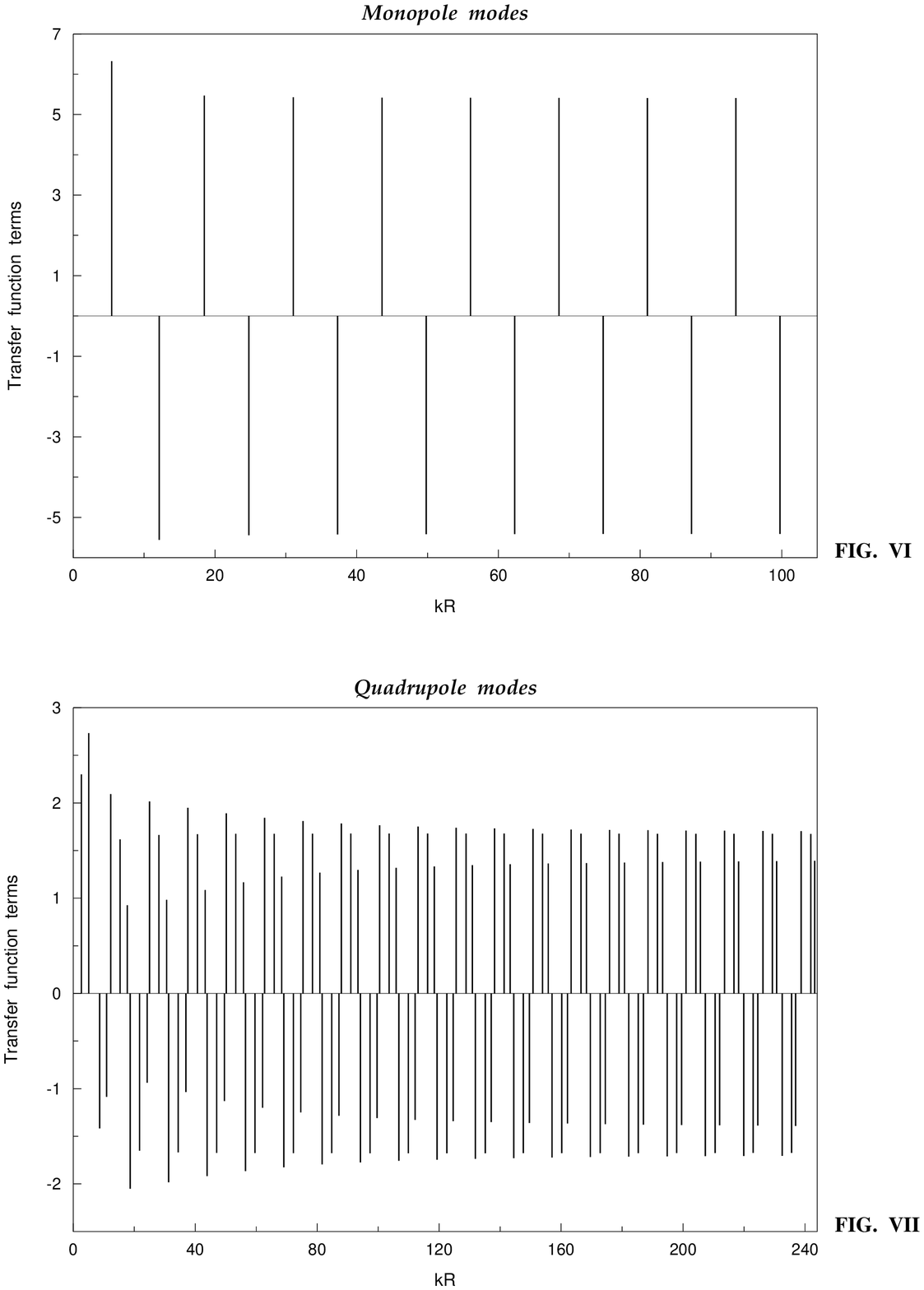,width=16.2cm}
\end{figure}

\end{document}